\documentclass{kapproc} % Computer Modern font calls
\usepackage{epsfig}
\begin{document}

\articletitle[A Sample contribution]
{Guiding the Way to Gamma-Ray Sources: X-ray Studies of Supernova Remnants}

\author{Patrick Slane}
\affil{Harvard-Smithsonian Center for Astrophysics\\
60 Garden Street\\ 
Cambridge, MA 02138\\
USA}
\email{slane@cfa.harvard.edu}

%\author{Second Author}
%\affil{Author Affiliation\\
%Second Line of Affiliation}
%\email{secondauthor@anotheruniv.edu}

\begin{keywords}
Supernova Remnants, $\gamma$-ray sources
\end{keywords}

\begin{abstract}
Supernova remnants have long been suggested as a class of potential
counterparts to unidentified $\gamma$-ray sources. The mechanisms
by which such $\gamma$-rays can arise may include emission from a
pulsar associated with a remnant, or a variety of processes
associated with energetic particles accelerated by the SNR shock.
Imaging and spectral observations in the X-ray band can be used to
identify properties of the remnants that lead to $\gamma$-ray emission,
including the presence of pulsar-driven nebulae, nonthermal X-ray
emission from the SNR shells, and the interaction of SNRs
with dense surrounding material. 

\end{abstract}

\section{Introduction}
Supernova remnants and their associated pulsars represent an energetic
class of objects which is intimately connected to sources of $\gamma$-ray
emission in our Galaxy. The connection between SNRs and the energetic cosmic
rays that pervade the Galaxy has long been assumed, for example; 
shock acceleration
by the SNR blast wave provides ample energy for the production of 
multi-TeV particles, and the presence of nearby material in dense clouds
from which the remnant progenitors collapsed forms a natural target
for pion production with subsequent $\pi^0 \rightarrow 2\gamma$ decay. 
Nonthermal bremsstrahlung of electrons off ambient material, as well as
inverse Compton scattering of electrons off ambient photons, can also lead to
$\gamma$-ray production -- and in many cases are the dominant mechanisms.
SNRs are thus strong candidates for the emission
of $\gamma$-rays. Recent observations have revealed remnants
whose emission is dominated by nonthermal X-rays from these energetic
particles, providing direct evidence of TeV electrons, and leading to 
the detection of two remnants as VHE $\gamma$-ray sources.

Gamma-ray studies to date have clearly identified pulsars as strong
candidates for $\gamma$-ray emission (see Thompson -- these proceedings), 
and SNRs hold another
key in this regard. Recent X-ray observations (e.g. Slane et al. 1997; 
Harrus, Hughes, \& Slane 1998; Harrus \& Slane 1999) have revealed numerous
new synchrotron nebulae in SNRs which, to the best of our understanding,
must house magnetospherically active neutron stars. Studies with 
{\it Chandra} and XMM-Newton can
realistically be expected to expand much of the ground breaking work
done with ASCA in this regard, thus improving our ability to identify
promising candidates for energetic $\gamma$-ray production. Here I
review recent and ongoing observational X-ray work on 
composite remnants, SNR interactions with molecular clouds, and nonthermal
emission from shell-type SNRs, and discuss how these might act as guides
to sources of $\gamma$-ray emission.

\section{Gamma-Rays From SNRs}

The large positional uncertainties for EGRET sources, coupled with the extended
nature of SNRs, make the clear identification of direct associations between
the two populations a difficult task (Figure 1, left). Moreover, statistical
assessment of the likelihood of a given association is complicated by the fact
that the distribution of potential source candidates (e.g. SNRs, pulsars,
OB associations) are not unrelated
(Kaaret \& Cottam 1996; Romero et al. 1999; Grenier, 2000).
These results render positional coincidences uncompelling as the only
evidence for an association between a particular SNR and EGRET source.
X-ray observations probe several
characteristics of SNRs and their environments which are directly related
to $\gamma$-ray production, and are thus of considerable importance
in establishing such associations. In addition, such observations can
provide candidate SNRs for further, more sensitive $\gamma$-ray observations.

There are three primary means by which $\gamma$-rays might be associated
with an SNR. Though only indirectly connected, the most prevalent
is emission directly from an associated pulsar and/or a pulsar-driven
synchrotron nebula. This is
likely to be a viable mechanism for only somewhat young pulsars,
and it is thus possible that the associated SNRs will still be
detectable. A more direct mechanism is $\pi^0$ decay through interaction
of the SNR shell with dense material in a nearby molecular cloud.
Finally, nonlinear shock acceleration at the SNR shell may result in
inverse Compton and/or nonthermal bremsstrahlung radiation from energetic 
electrons accelerated by the SNR. Each of these scenarios can be associated
with particular characteristics of the SNR X-ray emission. I discuss 
these cases briefly, and present possible examples of each.

\subsection{Pulsars and Synchrotron Nebulae}

%%%%%%%%%%%%%%%% Figure 1 %%%%%%%%%%%%%%%%%%%%%%%%%%%%%%%%%%%%%%%%%
\begin{figure}[t]
\centerline{
\epsfig{file=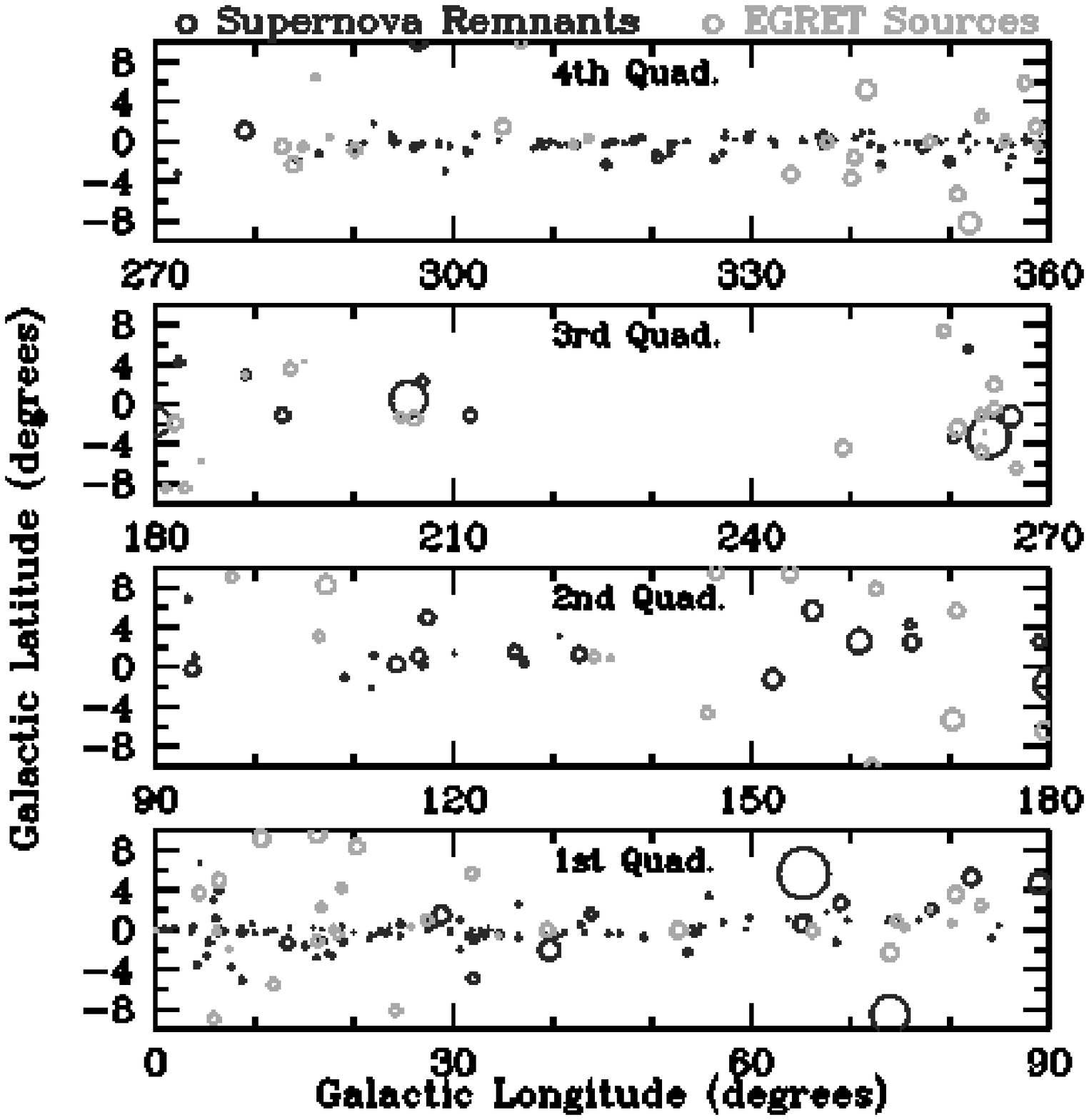,width=2.5in}
\epsfig{file=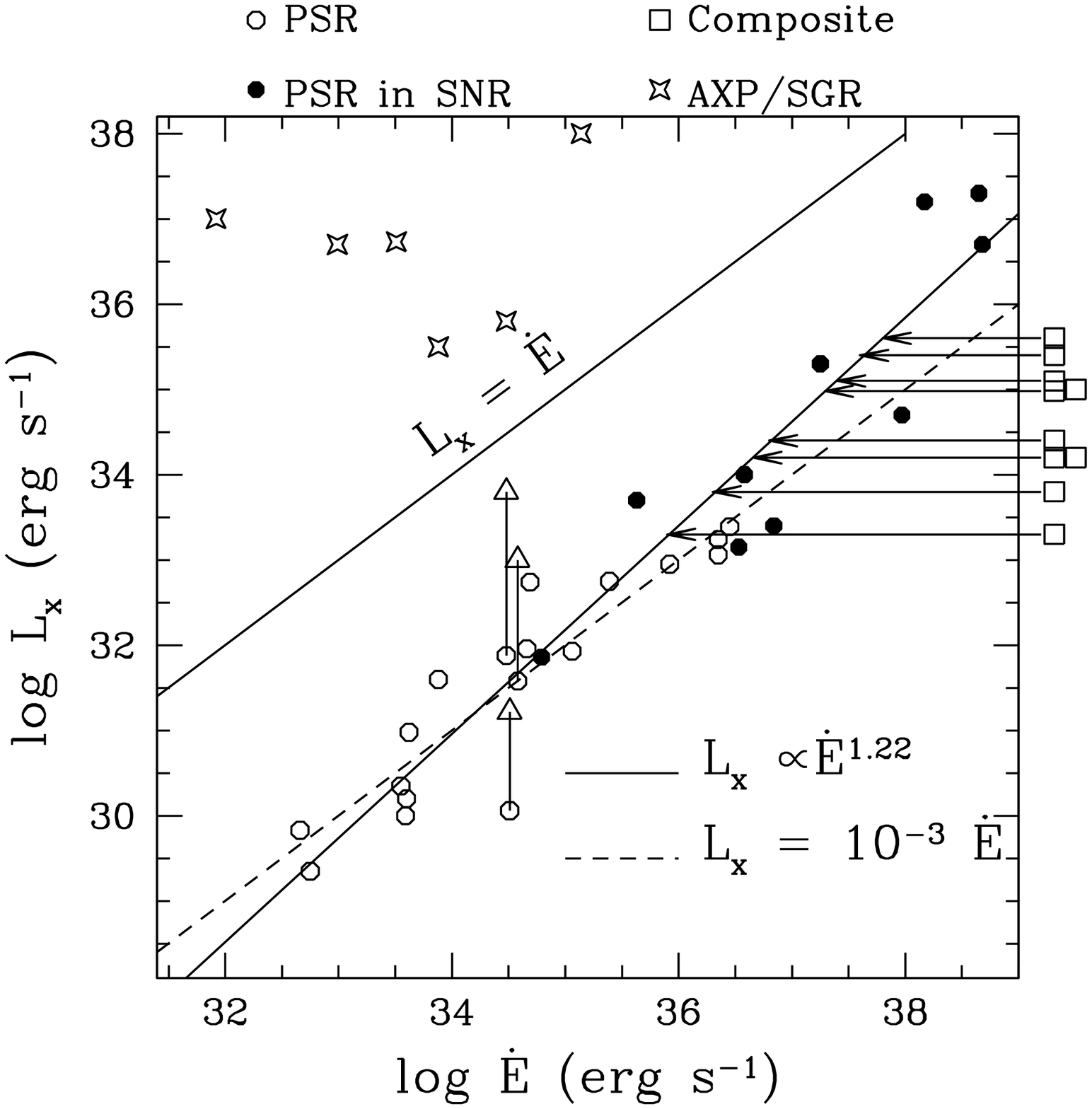,width=2.5in}
}
\caption{Left: Distribution of SNRs (black) and EGRET sources (grey) in
the Galactic plane. Circles correspond approximately to SNR size or EGRET
position error circles. Right: $L_x$ vs. $\dot E$ for pulsars and plerions.
Triangles correspond to pulsars with identified cooling components, and the
associated circles have this component subtracted. Luminosities include
emission from associated synchrotron nebula where known.
For composite SNRs whose pulsars are not detected, $L_x$ is plotted along
the right axis. Arrows indicate where these objects would reside on the
$L_x$ vs. $\dot E$ diagram. Note the good agreement between these values
and those for young pulsars in SNRs (filled circles).}
\end{figure}
%%%%%%%%%%%%%%%%%%%%%%%%%%%%%%%%%%%%%%%%%%%%%%%%%%%%%%%%%%%%%%%%%%%%

Since the majority of SNe result from massive star collapse, and
most such collapses are expected to form neutron stars (NSs), it is reasonable
to expect that young NSs should be associated with most young SNRs.
If all young NSs are rapidly rotating, magnetospherically
active pulsars, the corollary would then suggest that these SNRs
are likely to be sources of energetic $\gamma$-ray emission since
pulsars are a well-established class of $\gamma$-ray sources.
In practice, however, there are numerous examples of young SNRs for which
no associated NS is observed. Moreover, recent observations
have revealed compact X-ray sources, apparently associated with
SNRs, whose properties differ considerably from those of Crab-like
pulsars (see Caraveo, Bignami, \& Tr\"umper 1996; Kaspi 2000). These include 
the Anomalous X-ray Pulsars (AXPs) such as 1E~2259+586 in the remnant CTB 109, 
the radio-quiet neutron stars such as PKS~1209$-$51/52 in Puppis~A
(Zavlin, Tru\"umper, \& Pavlov 1999), 
and the point source in Cas A, whose properties
are unlike any of the above (Pavlov et al. 2000).
In cases where an X-ray emitting synchrotron nebula (or ``plerion'') is
observed, however, we infer the presence of a young pulsar powering the nebula.
Remnants displaying such a component are thus good candidates for $\gamma$-ray
emission.

Empirically, it is found that there exists a relationship between $\dot E$,
the pulsar energy loss rate, and $L_x$, the total X-ray luminosity from the
pulsar (not including any cooling component) and its associated synchrotron
nebula (Seward \& Wang 1988; Becker \& Tr\"umper 1997). 
Thus, even in the absence
of direct detection of a pulsar, its properties can be inferred from the
nonthermal X-ray spectrum of the plerion.
This is illustrated in Figure 1 (right) where I have plotted values for known
pulsars. Synchrotron nebulae in composite SNRs for which no pulsar has yet been
identified are shown as well.
It is important to note, however, that variations in the $L_x$~vs.~$\dot E$
relation are large. This is due, at least in part, to the variety of conditions
inherent to the systems used to establish the relation (bow shock nebulae,
plerions interacting with supernova ejecta, etc.). This has been addressed
in recent work by Chevalier (2000), who has 
developed a one-zone model for the X-ray properties of pulsar nebulae which
accounts for differences in the relative efficiency with which $L_x$
is produced, based on the cooling regime in which the electrons find
themselves after passing through the wind termination shock. 

\subsubsection{G21.5$-$0.9}
Although we infer the presence of an active, young NS with
any X-ray synchrotron nebula, it is by no means the case that pulsars
have been identified for each. G21.5$-$0.9, for example, is a well-defined
plerion for which radio observations show no direct
evidence of a compact central
source. {\it Chandra} observations (Slane et al. 2000) reveal a compact
source at the center of the nebula, but the high resolution images
clearly demonstrate that the source is slightly extended.
Using the radio spectrum to estimate the magnetic field strength of
the nebula under the assumption of equipartition between the field
and the particles, and requiring that this associated magnetic pressure
act to confine a pulsar wind driven by the spin-down power, 
$\dot E = 3.5 \times 10^{37}d_5^{1.6}{\rm\ erg\ cm}^{-2}{\rm\ s}^{-1}$, of a
central pulsar (using the empirical $\dot E$ vs. $L_x$ relationship,
with $d_5$ as the distance in units of 5~kpc) indicates
that the extended size of the central emission is consistent with
the radius of the stand-off shock where the wind is confined.
A smaller $\dot E$ is predicted by the model of Chevalier (2000), resulting
in a smaller estimated shock radius, but one still in rough agreement
with the observations.
As the particle wind traverses the standoff shock, the nebular structure
changes from a particle-dominated to a magnetic-dominated flow (Kennel
\& Coroniti 1984). The finite lifetime of the synchrotron emitting
electrons results in a steepening of the spectrum with radius,
an effect clearly observed in G21.5$-$0.9 (Slane et al. 2000).
Thus, the X-ray observations indicate the presence of an energetic pulsar in
G21.5$-$0.9. The inferred value of $\dot E/d^2$ (a rough indicator of the
expected $\gamma$-ray flux) is larger than that for
many of the known $\gamma$-ray pulsars (or for only several, using the value
estimated by Chevalier) suggesting that sensitive $\gamma$-ray observations of
this source may be of interest.

%%%%%%%%%%%%%%%%%%%%%%%%% Figure 2 %%%%%%%%%%%%%%%%%%%%%%%%%%%%%%%%%
\begin{figure}[t]
\centerline{\epsfig{file=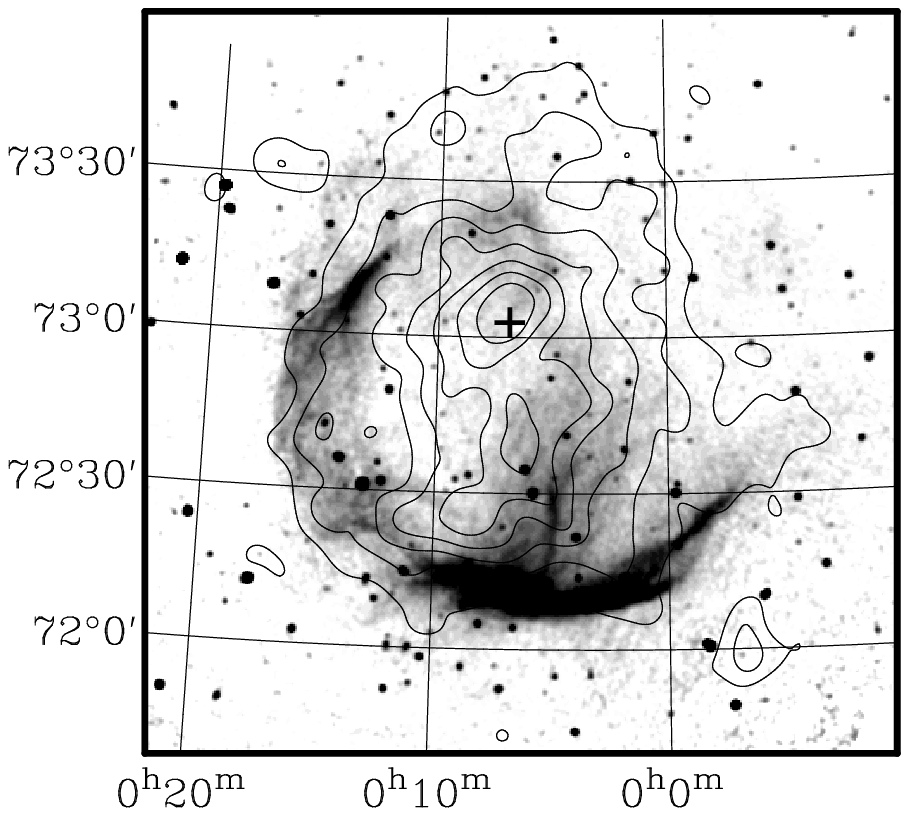, width=2.5in}
            \epsfig{file=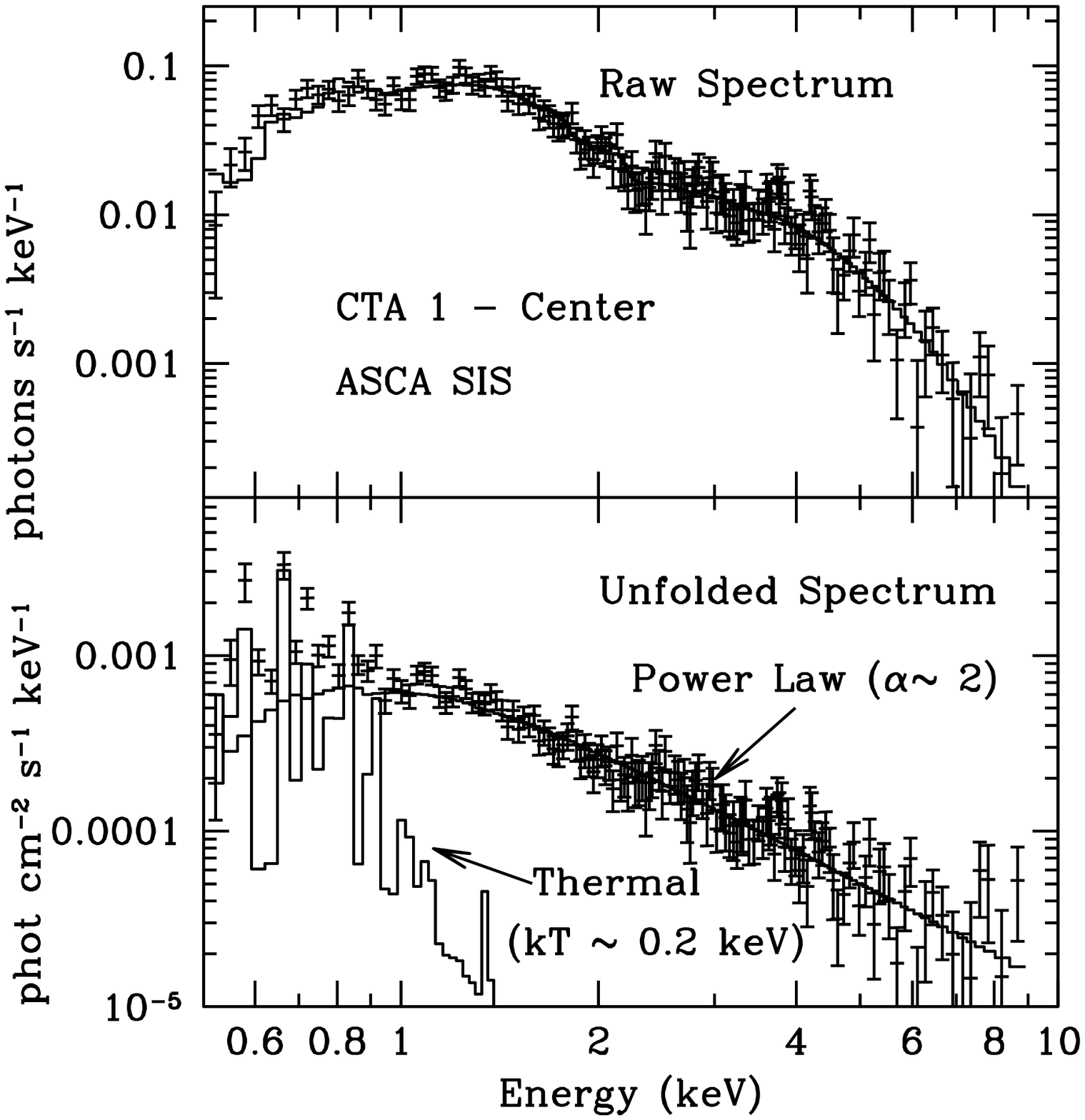, width=2.5in}}
\caption{Left: Continuum image of CTA~1 at 1420 MHz with X-ray
contours from the ROSAT PSPC.  (Radio image kindly provided by T. Landecker.)
Note the incomplete shell-like radio morphology contrasted with the centrally
bright X-ray morphology. The position of the faint point source J000702+7302.9
is indicated with a cross.
Right: ASCA data extracted from SIS chip centered on central
region of CTA 1. Upper: Raw spectrum with best-fit power-law plus
thermal model. Lower: Unfolded spectrum and model components.}
\end{figure}
%%%%%%%%%%%%%%%%%%%%%%%%%%%%%%%%%%%%%%%%%%%%%%%%%%%%%%%%%%%%%%%%%%%%

\subsubsection{CTA1 and 2EG~J0008+7303}

Another example of the pulsar-plerion connection of particular interest in the
context of $\gamma$-ray sources in that of CTA~1.
In the radio band, CTA~1 (G119.5+10.3) consists of an incomplete shell
(Figure 2)
with faint emission extending to the northwest in what may be a breakout
region (Pineault et al. 1997). The X-ray morphology is centrally brightened
with emission bounded by the radio shell in the south and west, and with faint
emission extending into the breakout region. ROSAT PSPC observations indicate
a two-component spectrum for which the hard component dominates in the central
regions but becomes weaker in the outer regions of the remnant (Seward,
Schmidt, \& Slane 1995). A faint X-ray source (J000702+7302.9) is found at
the peak of the central X-ray emission (Figure 2).

ASCA observations of CTA~1 (Slane et al. 1997) reveal a nonthermal central
spectrum clearly indicating the composite nature of this remnant (Figure 2).
This is bolstered by the presence of the EGRET source 2EG~J0008+7303 whose
error circle includes the position of the faint X-ray source (Brazier
et al. 1997) and whose spectrum, when extrapolated to the soft X-ray band,
is consistent with that measured for J000702+7302.9 (although the latter
fact is not necessarily expected from emission models).

%The X-ray data imply that CTA~1 has an age of $2 \times 10^{4}
%D_{1.4}$~yr has swept up $32.6 D_{1.4}^{5/2} M_\odot$ of material from a
%surrounding
%medium with $n_0 = 3.7 \times 10^{-2} D_{1.4}^{-1/2}{\rm\ cm}^{-3}$. The
%explosion energy is $1.5 \times 10^{50} D_{1.4}^{5/2}{\rm\ erg}$ and the
%central thermal pressure is $2.6 \times 10^{-11} D_{1.4}^{-1/2}
%{\rm\ dyne\ cm}^{-2}$. Assuming pressure equilibrium, this implies a plerionic
%magnetic field of $25 D_{1.4}^{-1/4}{\ \mu\rm G}$.
%The magnetic energy density is roughly $10^4$ times that of the X-ray
%emitting electrons. However, accounting for synchrotron losses and adiabatic
%expansion of the nebula, Slane et al. (1997) find approximate 
%equipartition between particle
%and magnetic flux at the time of injection from the central engine.

The nonthermal X-ray luminosity is $5.6 \times 10^{33} D_{1.4}^{2}{\rm\ erg\
s}^{-1}$ leading to an inferred pulsar energy loss rate of $\dot E = 1.7
\times 10^{36} D_{1.4}^{1.4}{\rm\ erg\ s}^{-1}$.
If correct, the value of $\dot E/d^2$ 
is exceeded by only 4 pulsars, all of which (along with 4-5 with
smaller values) are known $\gamma$-ray sources. The X-ray measurements thus
strongly indicate that 2EG~J0008+7303 is associated with the X-ray source and
synchrotron nebula in CTA~1. Modeling of outer gap $\gamma$-ray production
(Zhang \& Cheng 1998) indicates that a $\gamma$-ray pulsar scenario for
this source is indeed feasible.
Deeper X-ray observations of this source are
currently underway with XMM-Newton.

\subsubsection{Other Plerionic Candidates}
In addition to the Crab and Vela pulsars, both associated with synchrotron 
nebulae, and both well-known $\gamma$-ray pulsars, there are several EGRET
sources for which X-ray observations suggest a pulsar (or pulsar-driven
nebula) counterpart. Halpern et al. (2001) use X-ray, radio, and optical
data to show that 3EG~J2227+6122 may be associated with a pulsar-powered
source.  An incomplete circular shell within the EGRET error circle 
is seen in VLA sky survey data, and ASCA data from this region reveals 
emission well-described by a power law. The radio image resembles a
bow-shock nebula or wind-blown bubble, and the X-ray properties are
consistent with such an interpretation.

ASCA observations of the region around 2EG~J1418$-$6049 (Roberts and
Romani 1998) also suggest a pulsar wind nebula interpretation for this
source. The X-ray image reveals an extended nonthermal object whose
properties are consistent with being a nebula energized by a pulsar
with a characteristic age of $\sim 10^{4.5}$~yr.

Extended nonthermal X-ray emission is also observed within the error
circle of 2EG~J1811$-$2339 (Oka et al. 1999). These authors suggest
that the source may be a synchrotron nebula associated with a currently
unknown pulsar, and that the nebula may be interacting with a dense
cloud of material seen in CO observations. In this case, the $\gamma$-ray
emission may only be indirectly associated with the pulsar, which may act
as a source of electrons that undergo shock acceleration at the cloud
interaction site.
Several other SNRs with nearby pulsars have been suggested as counterparts for
EGRET sources (e.g. W44 and W28), but it appears that the $\gamma$-ray
emission from these sources is also associated with the remnants and not
the pulsars. Such interaction processes are discussed in more detail below.

\subsection{Particle Acceleration by SNR Shocks}

Beyond the early $\gamma$-ray emission fueled by the
radioactive decay of unstable isotopes formed in supernovae,
the mechanical energy released in the explosions can be
transformed into $\gamma$-ray emission over a longer period of
time. As the blast wave from the explosion expands, material from
the surrounding circumstellar material (CSM) and ISM is swept up
into a shell of hot gas. For an ideal $\gamma = 5/3$ gas, the
shock jump conditions yield an increase in density by a factor of
4 and a postshock temperature 
$$T = \frac{3 \mu m}{16 k} V_s^2$$
where $V_s$ is the shock speed, $m$ is the proton mass, and $\mu$ is 
the mean molecular weight of the gas ($\mu \approx 0.6$).
This shock-heated gas yields the familiar X-ray emission, characterized 
by a thermal bremsstrahlung spectrum accompanied by strong emission lines.

As increasing amounts of material are swept up by the blast wave, the 
shock is decelerated and the ejecta from the explosion encounter the 
dense shell. A reverse-shock is generated, heating the ejecta. At early
times, then, the X-ray spectrum is dominated by emission from the ejecta.
As the amount of swept-up material increases, the spectrum becomes 
dominated by emission from material with ISM abundances.

In addition to thermal heating of the swept-up gas, some fraction of the
shock energy density goes into nonthermal production of
relativistic particles through diffuse shock acceleration. 
Particle acceleration by SNR shocks has, of course, been suggested for
decades as a process by which cosmic rays are produced. Radio observations
of SNRs provide ample evidence of electrons with GeV energies through
synchrotron radiation in the compressed magnetic field at the remnant
shell. At higher particle energies, $\gamma$-ray production may
result from nonthermal bremsstrahlung of electrons colliding with
ambient gas, from inverse Compton scattering of ambient photons,
and from the decay of neutral pions created by the collision of
energetic protons. If the relativistic particle component of the
energy density becomes comparable to that of the thermal component,
the shock acceleration process can become highly nonlinear. The gas
becomes more compressible, which results in a higher density and
enhanced acceleration. This process has been considered in detail
by Baring et al. (1999) who present a model for the radio to
$\gamma$-ray emission. The relative contributions from the different
$\gamma$-ray production mechanisms depend highly on ambient conditions,
and X-ray studies of these SNRs reveal these conditions and can provide 
spectral measurements which highly constrain the models. Here I summarize
the X-ray properties of several SNRs which may be associated with
$\gamma$-ray emission, in an effort to illustrate these ideas.

\subsection{Pion Decay and Molecular Cloud Interactions}
If the blast wave of an SNR is capable of accelerating protons (or other
ions) to sufficiently high energies, then the interaction of these
particles with ambient material can result in the production of neutral
pions which subsequently decay into $\gamma$-rays. The $\gamma$-ray
flux for such a process was calculated by Drury, Aharonian, \& V\"olk
(1994). For a power-law particle spectrum with spectral index 
$\alpha \sim 2.1 - 2.3$, the flux is
$$ F_\gamma (\ge E_{\rm TeV}) \approx 5 \times 10^{-11} \epsilon E_{51} 
d_{\rm kpc}^{-2} n E_{\rm TeV}^{-\alpha + 1} {\rm\ cm}^{-2} {\rm\ s}^{-1}$$
where $\epsilon$ is the fraction of the blast wave energy ($E_{51}$, in
units of $10^{51}$~erg) converted into cosmic ray energy, 
$d_{\rm kpc}$ is the SNR distance in kpc,
$n$ is the number density of the ambient medium, and $E_{\rm TeV}$ is
the $\gamma$-ray energy in TeV. The leading numerical factor actually 
depends on $\alpha$; I have taken an average value which is accurate to
within a factor of two or so over the stated range for $\alpha$. This
spectrum is plotted in Figure 3 (left) for two values of $\epsilon$, the larger
of which is probably an overestimate of what can typically be extracted
from this process. Here I have assumed $n = 1 {\rm\ cm}^{-3}$, and
$E_{51} = d_{\rm kpc} = 1$. Also shown are approximate sensitivities for 
EGRET, GLAST, and VERITAS. Note that, for typical SNR parameters, the
predicted flux is near or below the EGRET sensitivity, although above
the sensitivity limits for GLAST or VERITAS. 

For cases in which the SNR is encountering dense material
(i.e. $n \gg 1$), such as in
an interaction with a nearby molecular cloud, the $\gamma$-ray flux
can be enhanced considerably. 
In a model for the evolution of SNRs in molecular
clouds, Chevalier (1999) argues that the $\gamma$-ray emission results
from the radiative shell of the remnant where some fraction of the
energy density in the postshock region goes into relativistic
particles. Bremsstrahlung interactions in the dense shell yield the
high energy emission, and synchrotron radiation from this population,
in the compressed magnetic field, can simultaneously produce
the observed radio emission.
Because massive stars live their lives quickly, they expire nearby the
molecular clouds from which they formed; the ensuing supernova explosion
can thus lead to direct interaction with the molecular clouds.
Such SNR/cloud interactions are most clearly
identified through OH maser emission (e.g. Frail et al. 1996),
although CO mapping also provides an important means of tracing
the molecular gas around SNRs. 

%%%%%%%%%%%%%%%%%%%%%%%% Figure 3 %%%%%%%%%%%%%%%%%%%%%%%%%%%%%%%%%%
\begin{figure}[t]
\centerline{
\epsfig{file=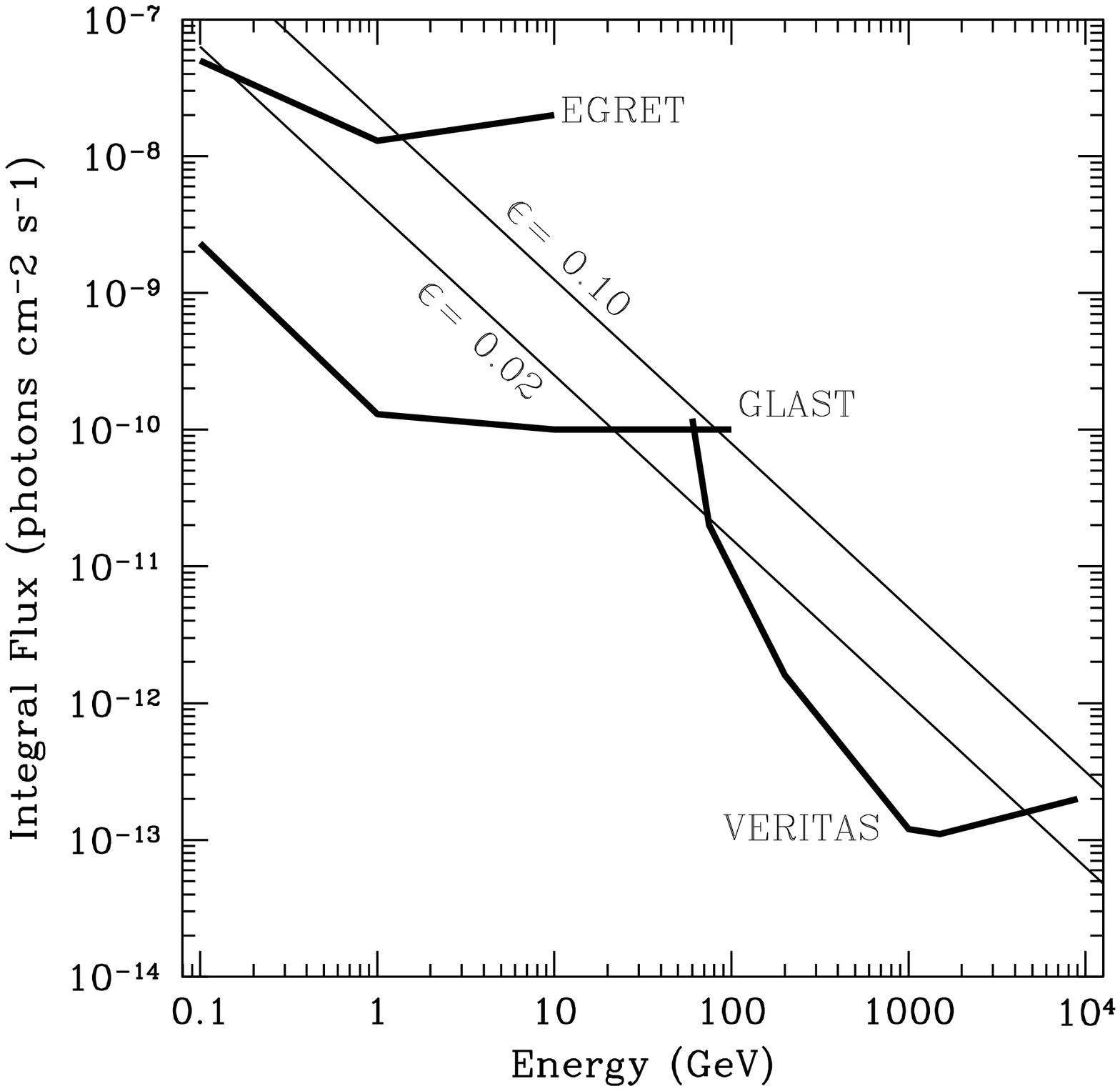,width=2.5in}
\epsfig{file=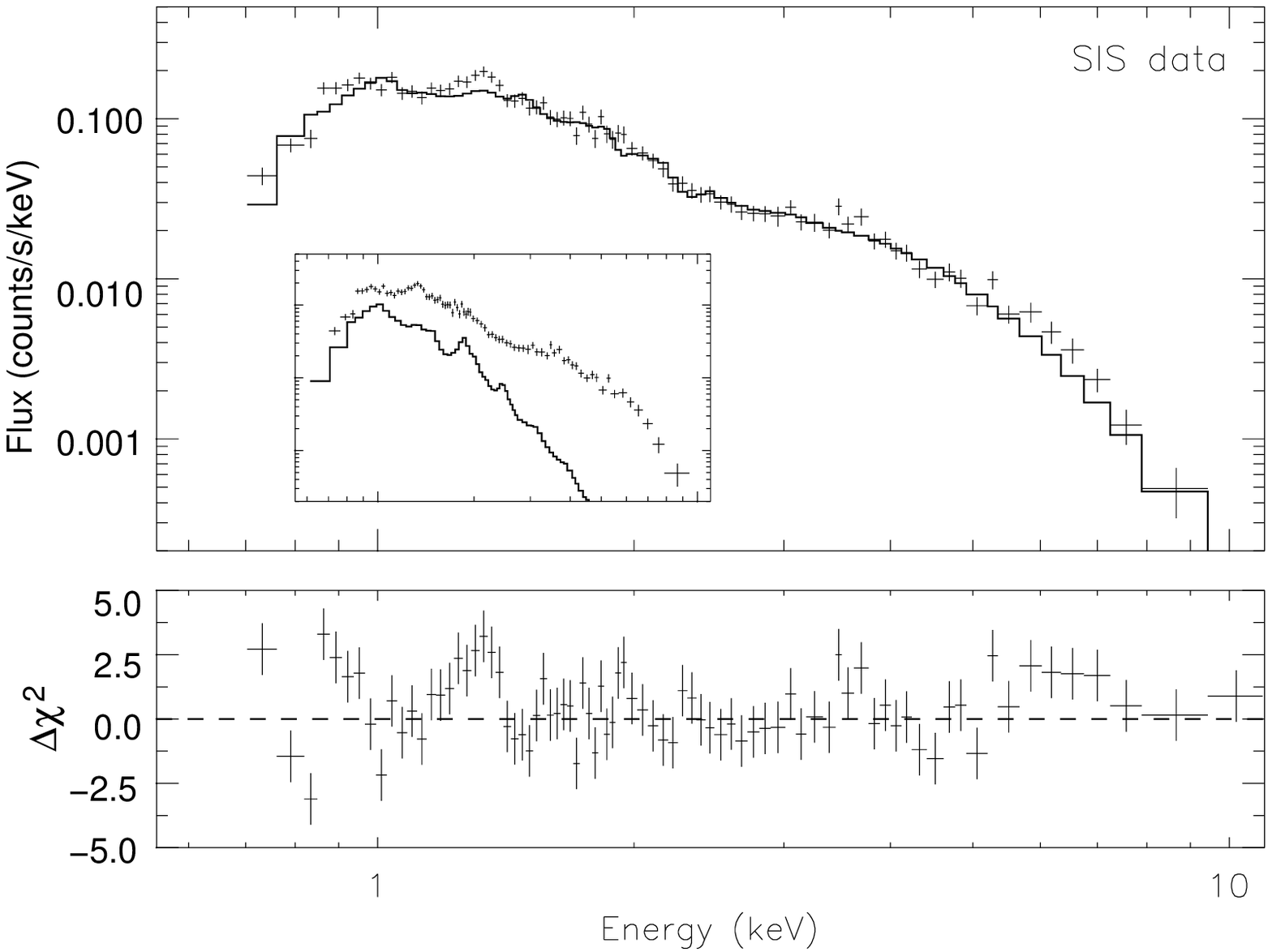,width=2.5in}
}
\caption{Left: Model pion-decay $\gamma$-ray flux from SNRs (see text).
Also shown are detection sensitivities for EGRET, GLAST, and VERITAS.
Right: X-ray spectrum from MSH~11-6{\sl 2} showing clear evidence of a
synchrotron component.
}
\end{figure}
%%%%%%%%%%%%%%%%%%%%%%%%%%%%%%%%%%%%%%%%%%%%%%%%%%%%%%%%%%%%%%%%%%%%

The X-ray properties of SNRs can also provide an indication of such 
interactions. SNR/cloud interactions can results in a significant 
brightening of the interaction region due to the increase in density.
For example, molecular clouds may be associated with
the so-called ``mixed morphology'' (MM) class of
remnants (Rho \& Petre 1998). These SNRs are characterized by a center-filled
X-ray morphology, as might be expected for a shell-type remnant whose 
central region contains a pulsar-powered nebula, but in this case
the emission is thermal in nature, and the radial temperature profile
is relatively flat. Several models have been put forth to explain
the brightness and temperature profiles for these SNRs. The cloud
evaporation model derived by White and Long (1991) has been applied
to the emission from 
W44 (Harrus et al. 1996), G272.2$-$3.2 (Harrus
et al. 2001), MSH~11$-$6{\sl 1A} (Slane et al. 2001), and others.
In this model the SNR evolves in a two-phase interstellar medium
(ISM) consisting of
cool, dense clouds and a low density intercloud medium. The SNR shock
sweeps past the clouds quickly, leaving them to slowly evaporate
in the hot SNR interior, thus contributing X-ray emitting material
to the central regions. Harrus et al. (1996) also considered a
model for W44 in which the remnant had entered the radiative phase,
thus leaving the shell too cool to radiate significantly in X-rays,
but allowing the emission from the hotter interior to persist. W44
is known to be interacting with an adjacent molecular cloud, which
is consistent with deceleration of the shock and evolution into
the radiative phase. Shelton
et al. (1999) extended this picture significantly by including the
effects of thermal conduction on smoothing out the temperature and
density profile of the X-ray emission material. Slane et al. (2001)
applied both models to emission from MSH~11$-$6{\sl 1A}. 

Other SNRs which fit into this MM class include W28 (Long et al. 1991)
and IC~443 (Petre et al. 1988, Asaoka \& Aschenbach 1994), two
remnants which (along with W44) have been suggested as counterparts
to EGRET sources; each is known to be interacting with a molecular
cloud. Romero et al. (1999) have compiled a list of SNRs with positional
coincidences with EGRET sources, and at least one-third of these
show some indication of the presence of molecular clouds. As noted
above, these positional coincidences are far from convincing evidence
that each of these SNRs are associated with $\gamma$-ray sources, but the
presence of dense material provides some supporting evidence. 

%%%%%%%%%%%%%%%%%%%%%%%% Figure 4 %%%%%%%%%%%%%%%%%%%%%%%%%%%%%%%%%%
\begin{figure}[t]
\centerline{\epsfig{file=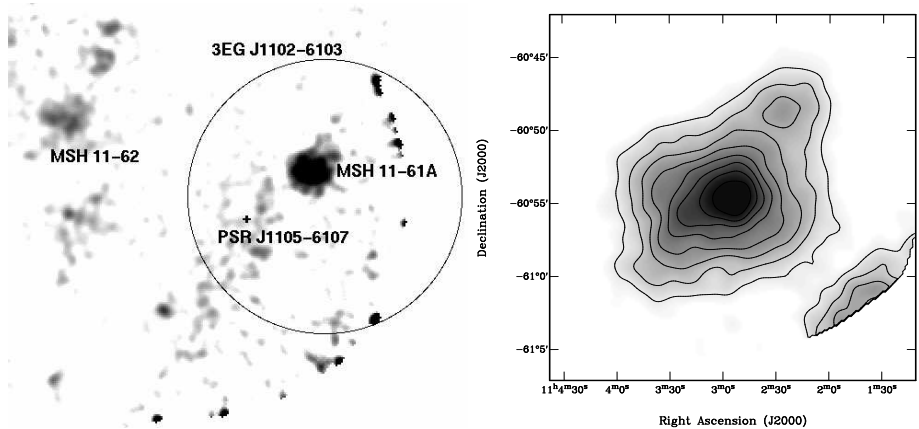,width=2.4in}
\epsfig{file=slane_fig4b.epsi,width=2.3in}}
\caption{
Left: X-ray emission from region surrounding 3EG~J1102$-$6103.
Right: ASCA GIS image of MSH~11--6{\sl 1A}. Emission at the
southwest edge of the field is associated with an off-axis source just
outside the primary telescope field of view.
}
\end{figure}
%%%%%%%%%%%%%%%%%%%%%%%%%%%%%%%%%%%%%%%%%%%%%%%%%%%%%%%%%%%%%%%%%%%%

\subsubsection{MSH~11$-$6{\sl 1A} and 3EG J1101$-$6103}
The unidentified EGRET source 3EG~J1101$-$6103 provides an
interesting case for which each of the topics above deserves
consideration.  Early catalogs associated this source
with MSH~11$-$6{\sl 2}. X-ray observations (Harrus, Hughes, \&
Slane 1998) reveal the remnant to be composite in nature, with
a two-component spectrum (Figure 3, right) consisting of a central
nonthermal component and an extended thermal component. As noted
above, the presence of a synchrotron nebula is indicative of a
young pulsar, making the association a reasonable one. An improved
position for the EGRET source revealed that MSH~11$-$6{\sl 2} falls
outside the error circle, however (Figure 4, left).  

Kaspi et al. (1997) identified a young pulsar PSR~J1105$-$6107
which lies within the error circle, and suggested an association
between the pulsar and the nearby remnant MSH~11$-$6{\sl 1A} (Figure
4). The value of $\dot E/d^2$ for this pulsar is quite low for
a $\gamma$-ray source, however, though higher than that for
PSR~B1055$-$52 which is detected by EGRET. Moreover, X-ray studies
indicate that an association between the pulsar and the SNR is
unlikely (Slane et al. 2001). 

It is of interest that MSH~11$-$6{\sl 1A} lies along the line of
sight to a dense molecular cloud whose distance ($\sim 7$~kpc), from 
CO measurements
appears consistent with that estimated for the remnant. Petruk (2000)
has suggested that the mixed-morphology nature of this remnant (and
others) is the result of enhanced emission from a cloud interaction
directly along the line of sight. Further, this same picture is
used to estimate the $\pi^0$-decay $\gamma$-ray luminosity. For a
cloud density of $\sim 40 {\rm\ cm}^{-2}$, the predicted flux (assuming
$\epsilon = 0.1$) matches that of 3EG J1101$-$6103. Efforts to 
model the observed X-ray temperature and brightness profiles under
this scenario are underway. An important point, however, is that
more complete modeling of the particle acceleration process in
shock-cloud interactions indicates that emission from pion decay 
falls considerably below that from bremsstrahlung (Gaisser et al. 1998,
Bykov et al. 2000).

\subsection{Nonthermal X-ray Emission from SNRs}

While the shock-heated ejecta and CSM/ISM components of shell-type SNRs
produce line-dominated X-ray spectra, as described above, a growing number
of remnants also reveal evidence of hard, nonthermal X-ray emission
components, apparently associated with high energy electrons accelerated
by the SNR shock.
In three cases -- SN~1006 (Koyama et al. 1995), G347.3$-$0.5 (Koyama et al.
1997; Slane et al. 1999), and G266.2$-$1.2 (Slane et al. 2001) -- the
nonthermal emission components completely
dominate the thermal components, and the X-ray spectra from the shells
are featureless.

Although the radio emission from SNRs provides direct evidence of GeV 
electrons, it is the recent detections of nonthermal X-rays from shell-type 
SNRs that provide direct evidence of particles accelerated to 
energies as high as
$10 - 100$~TeV, approaching the knee in the cosmic ray spectrum. Models
for nonlinear shock acceleration (e.g. Ellison, Berezhko, \& Baring 2000)
predict a strong correlation between the radio, X-ray,
and $\gamma$-ray flux that allow parameters of the acceleration
process to be derived.
It is thus of considerable interest to localize the regions of nonthermal
X-ray emission, and to investigate the broad-band spectra (particularly
in radio, X-rays, and $\gamma$-rays) from these regions.

%%%%%%%%%%%%%%%%%%%%%%%% Figure 5 %%%%%%%%%%%%%%%%%%%%%%%%%%%%%%%%%%
\begin{figure}[t]
\centerline{\epsfig{file=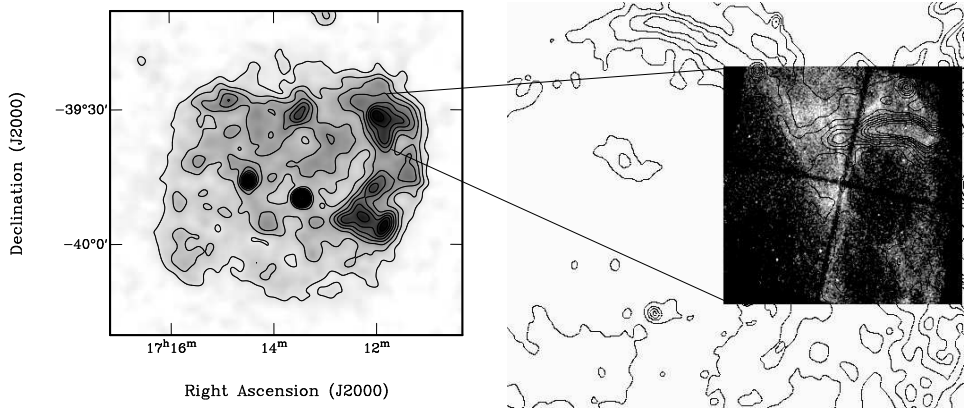,width=5in}}
\caption{Left: ROSAT PSPC image of G347.3$-$0.5. Right: {\em Chandra}
image of northwest rim, with radio contours from ATCA.}
\end{figure}
%%%%%%%%%%%%%%%%%%%%%%%%%%%%%%%%%%%%%%%%%%%%%%%%%%%%%%%%%%%%%%%%%%%%

In Figure 5, I present a preliminary {\it Chandra} image of the
X-ray emission from the NW limb of G347.3$-$0.5, whose
shell emission is nonthermal. Overlaid are radio contours
from the Australia Telescope Compact Array (ATCA), showing
good general correspondence between the components. This region of
the SNR has also been detected as a source of $\sim$TeV
$\gamma$-rays (Muraishi et al. 2000).
Preliminary modeling suggests a self-consistent picture in which the
X-ray and radio flux is synchrotron radiation from the
accelerated electrons, while the $\gamma$-ray flux represents
inverse-Compton scattering of energetic electrons off of the microwave
background (Ellison, Slane, \& Gaensler -- in preparation). The 
predicted GeV flux falls below the EGRET threshold, but should be
detectable with GLAST. Such measurements, along with more sensitive
X-ray observations, will provide significant information for constraining
the parameters of these models.

G266.2$-$1.2 was discovered by Aschenbach (1998) using
data from the ROSAT All-Sky Survey. Situated along the line of sight
to the Vela SNR, the X-ray emission stands out above the soft
thermal emission from Vela only at energies above $\sim 1$~keV.
ASCA observations show that the emission from G266.2$-$1.2 is
nonthermal (Slane et al. 2001), providing another example of
a shell-type SNR for which the X-ray emission is dominated 
by such processes. CO data (May et al. 1988) reveal a concentration of
giant molecular clouds -- the Vela Molecular Ridge -- at a distance
of $\sim 1 - 2$~kpc in the direction of Vela. The X-ray observations
of G266.2$-$1.2 indicated a column density consistent with the remnant
being located near, but probably not beyond, this molecular material.
More sensitive X-ray measurements are required to investigate the
shock structure and spectrum, for comparison with nonlinear shock
models. Observations at TeV energies are of considerable interest,
particularly given that SN~1006 and G347.3$-$0.5 have both been detected
at such energies. Given the relatively small distance to this remnant,
it would appear to be an excellent candidate for future observations
with GLAST as well.

%%%%%%%%%%%%%%%% Figure 6 %%%%%%%%%%%%%%%%%%%%%%%%%%%%%%%%%%%%%%%%%
\begin{figure}[t]
\centerline{
\epsfig{file=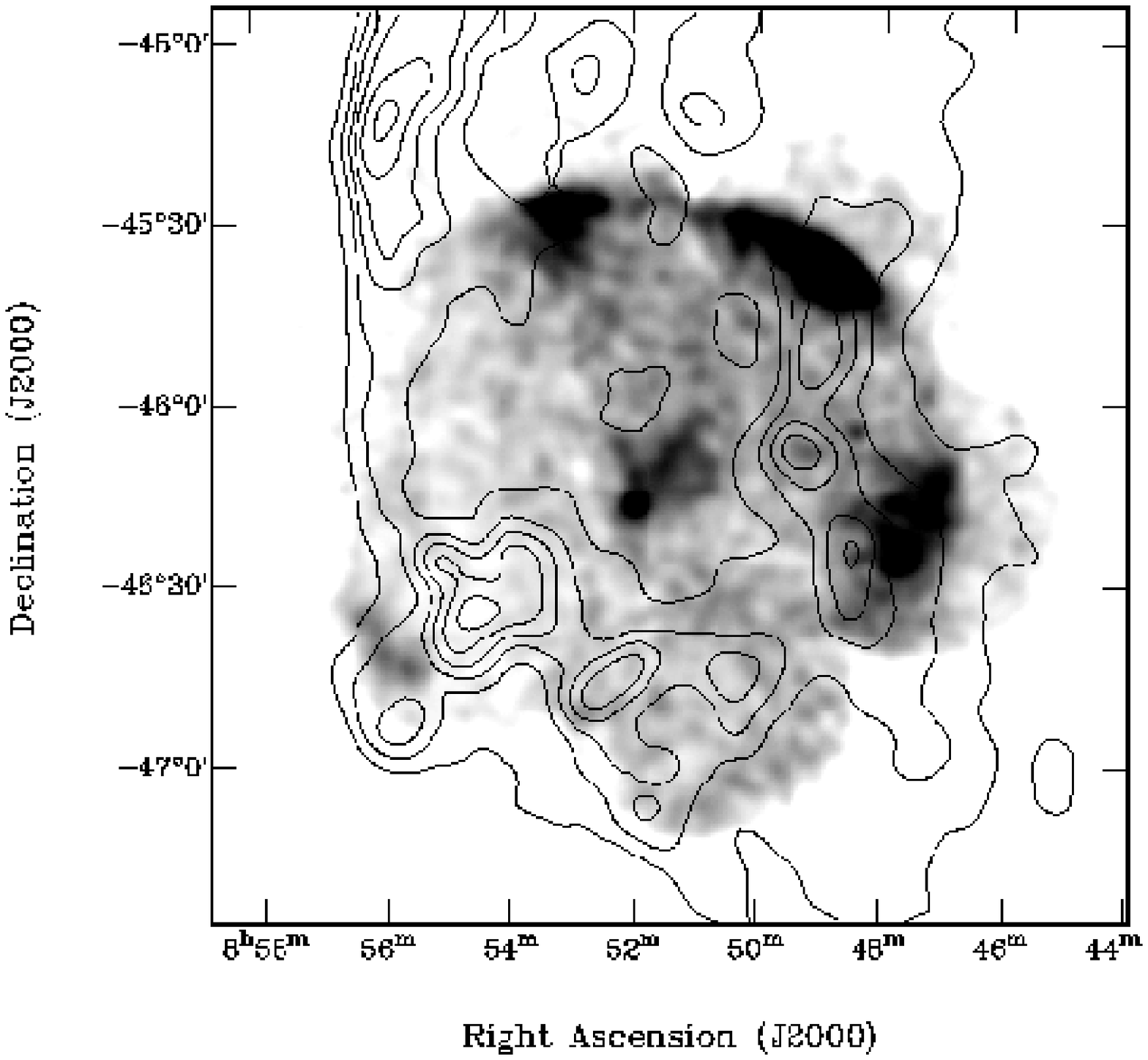,width=2.5in}
\epsfig{file=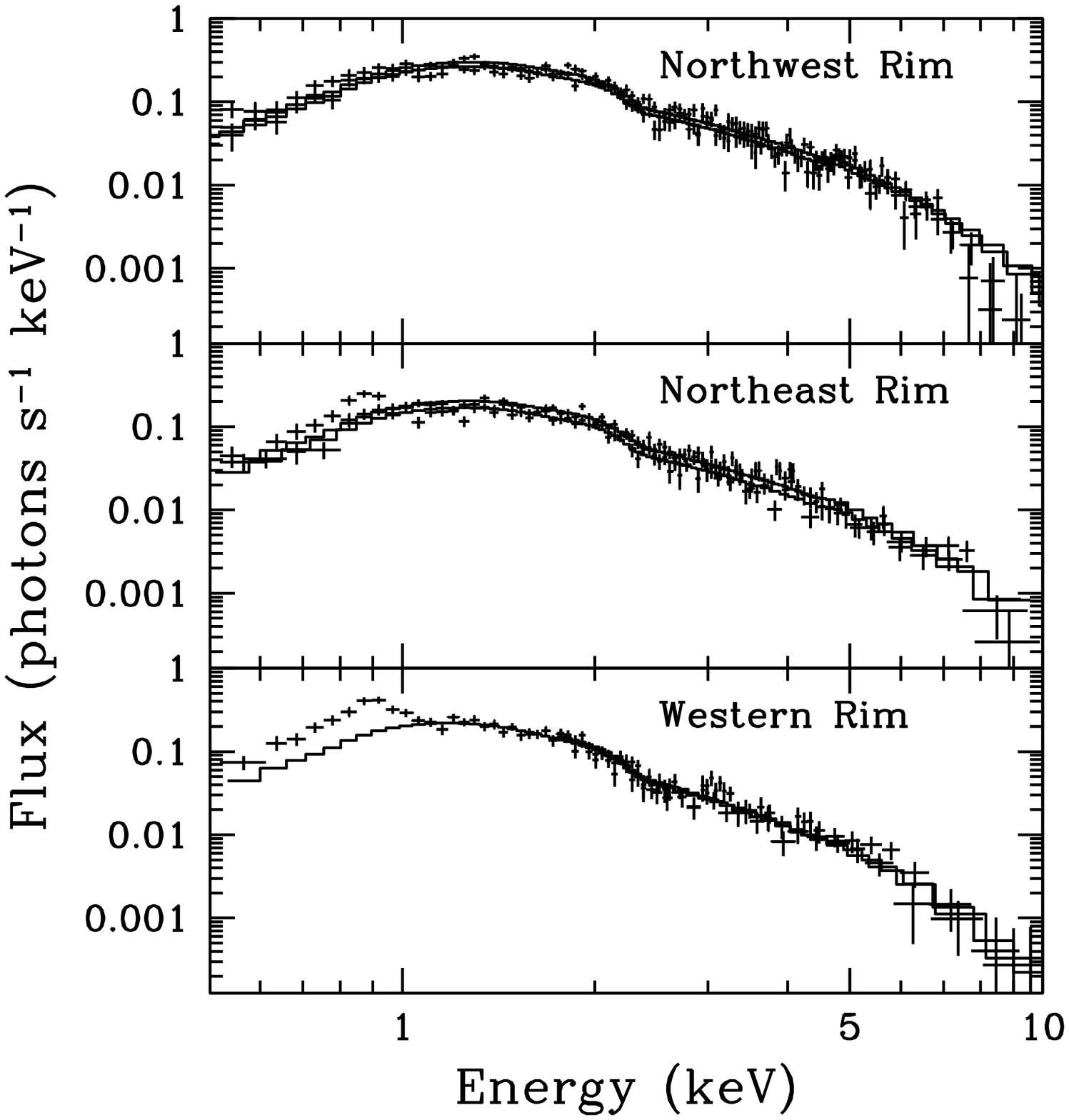,width=2.5in}}
\caption{Left: ASCA GIS image of G266.2--1.2 ($E = 0.7 - 10$~keV). The image
consists of a mosaic of 7 individual fields. Contours represent the outline
of the Vela SNR as seen in ROSAT survey data with the PSPC. Right:
ASCA spectra from both GIS detectors for regions of G266.2--1.2.
The featureless spectrum
is well described by a power law. Excess flux at low energies is
presumably associated with soft thermal emission from the Vela SNR.}
\end{figure}
%%%%%%%%%%%%%%%%%%%%%%%%%%%%%%%%%%%%%%%%%%%%%%%%%%%%%%%%%%%%%%%%%%%%

\section{Summary}

Several mechanisms exist by which SNRs may be associated with the emission
of energetic $\gamma$-rays. X-ray observations provide powerful
clues by which the properties of these mechanisms can be revealed.
Through the identification of pulsar-driven nebulae, nonthermal
emission from particles accelerated at the SNR shell, or sites of
interactions with dense clouds, X-ray studies of SNRs are helping
to resolve at least some of the questions centered on the nature of
unidentified EGRET sources. The prospects for continued success in
this area are extremely encouraging. The capabilities of current X-ray
observatories, coupled with upcoming $\gamma$-ray missions (both
space and ground based), promise not only to help identify the counterparts
to these sources of $\gamma$-rays, but also to provide new and
powerful constraints on developing models for such emission.

\begin{acknowledgments}
The author wishes to thank Don Ellison and Glenn Allen for helpful
discussions on nonthermal emission from SNRs. 
This work was supported in part by NASA through contract NAS8-39073
and grants NAG5-9106 and NAG5-9281.

\end{acknowledgments}


\begin{chapthebibliography}{1}

\bibitem{}
Asaoka, I. \& Aschenbach, B. 1994, {\it A\&A}, {\bf 284}, 573

\bibitem{}
Aschenbach, B. 1998 {\it Nature}, {\bf 396}, 141

\bibitem{}
Baring, M. G., Ellison, D. C., Reynolds, S. P., Grenier, I. A., \& 
Goret, P. 1999, {\it ApJ}, {\bf 513}, 311

\bibitem{}
Becker, W. \& Tr\"umper, J. 1997, {\it A\&A}, {\bf 326}, 682

\bibitem{}
Brazier, K.T.S., Reimer, O., Kanbach, G., \& Carraminana, A. 1997,
{\it MNRAS}, {\bf 295}, 819

\bibitem{}
Bykov, A. M., Chevalier, R. A., Ellison, D. C., \& Uvarov, Yu. A.
2000, {\it ApJ}, {\bf 538}, 203

\bibitem{}
Caraveo, P. A., Bignami, G. F., \& Tr\"umper, J. 1996, {\it A\&A Rev.}, 
{\bf 7}, 209 

\bibitem{}
Chevalier, R. A. 1999, {\it ApJ}, {\bf 511}, 798

\bibitem{}
Chevalier, R. A. 2000, {\it ApJ}, {\bf 539}, L45

\bibitem{}
Drury, L. O'C, Aharonian, F. A., \& V\"olk, H. J. 1994, {\it A\&A}, 
{\bf 297}, 959

\bibitem{}
Ellison, D. C., Berezhko, El G., \& Baring, M. G. 2000, {\it ApJ}, 
{\bf 540}, 292

\bibitem{}
Frail, D. A., Goss, W. M., Reynoso, E. M., Giacani, E. B., Green, A. J., 
Otrupcek, R. 1996, {\it AJ}, {\bf 111}, 1651

\bibitem{}
Gaisser, T. K., Protheroe, R. J., \& Stanev, T. 1998 {\it ApJ}, {\bf 492}, 21

\bibitem{}
Grenier, I. A. 2001, {\it A\&A} {\bf 364}, L93
\bibitem{}
Halpern, J. P., Gotthelf, E. V., Leighly, K. M., \& Helfand, D. J. 2001,
{\it ApJ}, in press

\bibitem{}
Harrus, I. M., Hughes, J. P., \& Helfand, D. J. 1996, {\it ApJ}, 
{\bf 464}, L161

\bibitem{}
Harrus, I. M., Hughes, J. P., \& Slane, P. O. 1998, {\it ApJ}, {\bf 499}, 273

\bibitem{}
Harrus, I. M. \& Slane, P. O. 1999, {\it ApJ}, {\bf 516}, 811

\bibitem{}
Harrus, I. M., Slane, P. O., Smith, R. K., \& Hughes, J. P. 2001, {\it ApJ} -- 
in press

\bibitem{}
Kaaret, P. \& Cottam, J. 1996, {\it ApJ}, {\bf 462}, L35

\bibitem{}
Kaspi, V. M., Bailes, M., Manchester, R. N., Stappers, B. W., Sandhu, J.
S., Navarro, J., \& D'Amico, N. 1997, {\it ApJ}, {\bf 485}, 820 

\bibitem{}
Kaspi, V. M. 2000, in ASP Conf. Ser. 202, Pulsar Astronomy 2000 and 
Beyond, ed.  M. Kramer, N. Wex, \& R. Wielebinski (San Francisco: ASP), 485

\bibitem{}
Kennel, C. F. \& Coroniti, F. V. 1984, {\it ApJ}, {\bf 283}, 710

\bibitem{}
Koyama, K. et al.  1995 {\it Nature} {\bf 378}, 255

\bibitem{}
Koyama, K. et al. 1997, {\it PASJ}, {\bf 49}, L7

\bibitem{}
Long, K. S., Blair, William P., Matsui, Yutaka, White, Richard L.
1991, {\it ApJ}, {\bf 373}, 567

\bibitem{}
May, J., Murphy, D. C., \& Thaddeus, P. 1988, {\it A\&ASS} {\bf 73}, 51

\bibitem{}
Muraishi, H. et al. 2000, {\it A\&A}, {\bf 354}, L57

\bibitem{}
Oka, T., Kawai, N., Naito, T., Horiuchi, T., Namiki, M., Saito, Y.,
Romani, R. W., \& Kifune, T. 1999, {\it ApJ}, {\bf 526}, 764

\bibitem{}
Pavlov, G. G., Zavlin, V. E., Aschenbach, B., Tr\"umper, J., \& Sanwal, D. 
2000, {\it ApJ}, {\bf 531}, L53

\bibitem{}
Petre, R., Szymkowiak, A. E., Seward, F. D., Willingale, R. 1988, 
{\it ApJ}, {\bf 335}, 215

\bibitem{}
Petruk, O, 2001, in Proceedings of NATO Advanced Study Institute 
``Astrophysical Sources of High Energy Particles \& Radiation'', (Erice, 
Italy, 11-21 November 2000, eds. J. P. Wefel, M. M. Shapiro, \&
T. Stanev.

\bibitem{}
Pineault, S., Landecker, T.L., Swerdlyk, C.M., \& Reich, W. 1997,
{\it A\&A}, {\bf 324}, 1152

\bibitem{}
Rho, J.-H. \& Petre, R. 1998, {\it ApJ}, {\bf 503}, L167

\bibitem{}
Roberts, M. S. E. \& Romani, R. W. 1998, {\it ApJ}, {\bf 496}, 827

\bibitem{}
Romero, G. E., Benaglia, P., \& Torres, D. F. 1999, {\it A\&A}, {\bf 348}, 868

\bibitem{}
Seward, F.D., Schmidt, B., and Slane, P. 1995,
{\it ApJ}, {\bf 453}, 284

\bibitem{}
Seward, F.D. \& Wang, Z.-R. 1988, {\it ApJ}, {\bf 332}, 199

\bibitem{}
Shelton, R. L., Cox, Donald P., Maciejewski, Witold, Smith,
Randall K., Plewa, Tomasz, Pawl, Andrew, \& Rózyczka, Michal
1999, {\it ApJ}, {\bf 524}, 192

\bibitem{}
Slane, P., Seward, F.D., Bandiera, R., Torii, K., \& Tsunemi, H. 1997,
{\it ApJ}, {\bf 481}, 225

\bibitem{}
Slane, P. et al. 1999, {\it ApJ}, {\bf 525}, 357

\bibitem{}
Slane, P., Chen, Y., Schulz, N. S., Seward, F. D., Hughes, J. P.,
\& Gaensler, B. M. 2000, {\it ApJ}, {\bf 533}, L29

\bibitem{}
Slane, P. et al. 2001, {\it ApJ} - accepted (see astro-ph/0010510)

\bibitem{}
Slane, P., Hughes, J. P., Smith, R. K., \& Petre, R. 2001, 
{\it ApJ} - submitted

\bibitem{}
White, R. L. \& Long, K. S. 1991, {\it ApJ}, {\bf 373}, 543

\bibitem{}
Zavlin, V. E., Tr\"umper, J., \& Pavlov, G. G. 1999, {\it ApJ}, {\bf 525}, 959

\bibitem{}
Zhang, L. \& Cheng, K. S. 1998, {\it A\&A}, {\bf 335}, 234

\end{chapthebibliography}
\end{document}